\documentclass[article,12pt,nofootinbib]{revtex4}
\usepackage[paper=letterpaper,margin=1.5in]{geometry}
\usepackage{epsfig,color}
\usepackage{graphicx}
\usepackage{amssymb,amsmath}
\usepackage{time}
\usepackage[varg]{txfonts}
\usepackage{hyperref}
\hypersetup{
  colorlinks,
  citecolor=green,
  linkcolor=blue}
\newcommand{\be}{\begin{equation}}
\newcommand{\ee}{\end{equation}}
\newcommand{\bc}{\begin{center}}
\newcommand{\ec}{\end{center}}

\renewcommand{\vec}[1]{\textnormal{\boldmath$#1$}}
\begin{document}

\bibliographystyle{revtex}

\begin{flushright}
{\normalsize
DESY-18-059\\
April 2018}
\end{flushright}

\vspace{.4cm}

\title
{Impedances of Anisotropic Round and Rectangular Chambers
}
\author{Igor Zagorodnov}
\affiliation{Deutsches Elektronen-Synchrotron, Notkestrasse 85,
22603 Hamburg, Germany}
\date{April 6, 2018}

\vspace{.4cm} 
\begin{abstract} 
We consider the calculation of electromagnetic fields generated by an electron bunch passing through an anisotropic transversally non-homogeneous vacuum chamber of round or rectangular cross-section with translational symmetry in the beam direction. The described algorithms are implemented in a numerical code and cross-checked on several examples.

\vfill \centerline
{Submitted to Physical Review Accelerators and Beams}
\end{abstract}

\maketitle


\section{Introduction}

Dielectric lined waveguides are under extensive study as accelerating structures excited by charged beams~\cite{Kan10}. Quartz  and  cordierite  structures  have  been  beam  tested,  and  accelerating  gradient  exceeding  100  MV/m  has  been  demonstrated~\cite{Gai09}. Several materials used for accelerating structures (sapphire, ceramic films etc) possess significant anisotropic properties. It is shown, for example, in~\cite{Sheinman} that the dielectric anisotropy causes a frequency shift in comparison to dielectric-lined waveguides with isotropic dielectric loadings.

The anisotropic dielectrics may be incorporated either intentionally or unintentionally (processing-induced anisotropy)~\cite{Yakovlev03}. Dielectric permittivity and conductivity depend on the direction of wave propagation and polarization in many materials. The anisotropy can have a significant effect on modal coupling and must be accounted for in the design and analysis of such structures.

There are many papers which describe impedance calculations of steady-state impedance for {\it isotropic} round and flat layered chambers~\cite{Mounet12, Piot12,Burov,Burov2,Ivanyan,Metral} with translational symmetry in the beam direction. The solutions for isotropic structures are obtained in analytical form or a field-matching approach can be used to reduce the problem to a simple matrix equation.  In this paper we consider {\it anisotropic} transversally non-homogeneous round and rectangular chambers where the field-matching technique does not work. 

We start in Section~\ref{sec:2} from formulation of the problem. Then we review the general form of the impedance for round and rectangular waveguides in non-relativistic case. For a special case of uniaxial anisotropy in the beam direction the same field matching approach as for the isotropic case~\cite{Mounet12} can be used. We consider shortly the required modifications in Section~\ref{sec:3}. For transversally non-homogeneous materials the structure can be approximated by many layers. However such approach could be computationally expensive as it requires  calculation of modified Bessel or exponential functions (of complex argument) for each layer. 

A fully anisotropic case is treated in Section~\ref{sec:4}. The field-matching technique does not work in this case. We consider several possible analytical formulations and choose one with non-singular differential equations. For this choice we describe a simple finite-difference scheme. With a proper permutation of mesh indexes we reduce a sparse matrix with 7 bands to a pentadiagonal one.  It allows a fast algorithm of complexity $O(N)$ ($N$-number of mesh steps) for calculation of impedances for non-homogeneous anisotropic materials. Open boundary conditions are formulated for the case when the last layer has an uniaxial anisotropy in the beam direction. 

The finite difference method allows treating of full anisotropy but could be time-consuming. For the case when the anisotropic layers are thin we suggest in Section~\ref{sec:5} a combination of the field matching and the finite-difference approaches.

Finally in Section~\ref{sec:6} the  described methods are cross-checked on several numerical examples. The algorithms are implemented in numerical code ECHO\cite{Zag16}.

%
\section{Problem formulation}\label{sec:2}
%

\begin{figure}[htbp]
\centering
\includegraphics*[height=60mm]{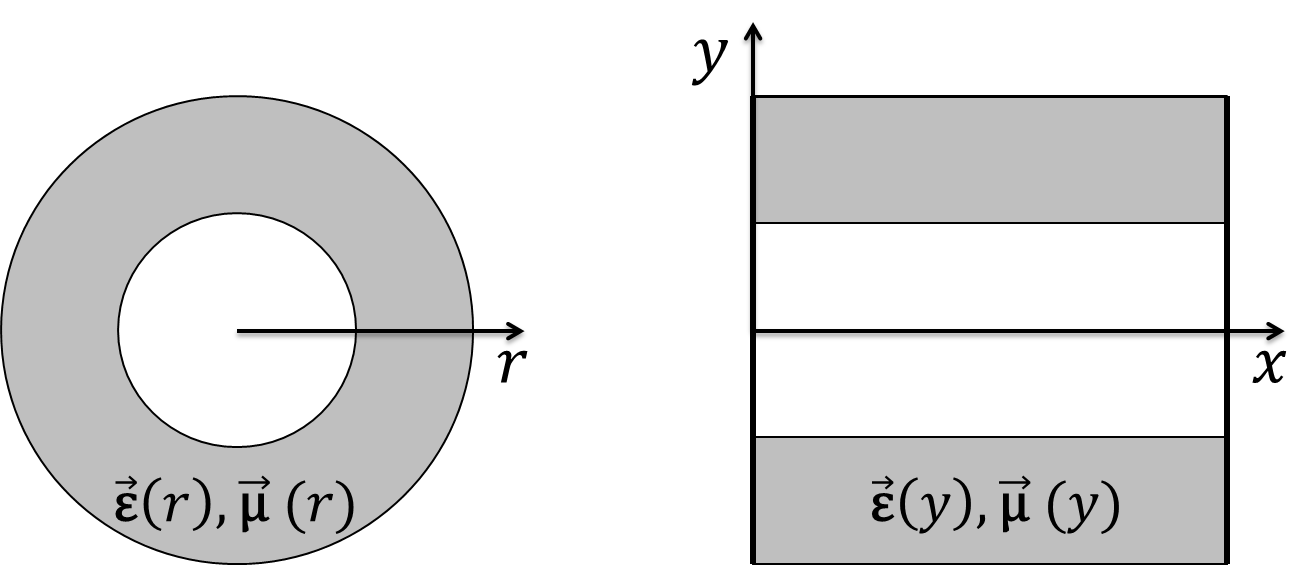}
\caption{Examples of "round" and "rectangular" geometry.}\label{Fig01}
\end{figure}

We consider a point-charge $q$ moving with constant velocity $v$ through a 
structure with round or rectangular cross-section. In the following we call the structure "round" if it is axially symmetric. If the structure has a constant width between two perfectly conducting planes and has rectangular cross-sections then we call such structure "rectangular". Fig.~\ref{Fig01} shows examples of round and rectangular structures. In the following we consider only an anisotropic materials with diagonal material permittivity and permeability tensors, where the optical axes coincide with coordinate ones. Hence their diagonals are given by complex vectors $\vec{\epsilon},\vec{\mu}$.

We assume that the charge is moving along a straight line parallel to the longitudinal axis of the system, and we neglect the influence of the wakefields on the charge motion. For round structures we will use cylindrical coordinates $r,\varphi,z$. The charge density in the frequency domain can be expanded in Fourier series
\begin{align}\label{EqChargeRound}
\rho(r,\varphi,z,k)=e^{-i k z/\beta}\sum\limits_{m=0}^\infty \rho_m(r)\cos(m (\varphi-\varphi_0)),\quad
\rho_m(r)=\frac{q\delta(r-r_0)}{\pi v r_0 (1+\delta_{m0})},
\end{align}
where $r_0,\varphi_0$ are coordinates of the point charge $q$, $\beta=v/c$, $c$ is velocity of light in vacuum, and $\delta_{m0}=1$ if $m=1$, 0 otherwise.

From the linearity of Maxwell's equations the components of the electromagnetic field can be represented by infinite sums:
\begin{align}\label{EqFieldsRound}
\begin{pmatrix}
H_{\varphi}(r,\varphi,z,k)\\E_r(r,\varphi,z,k)\\E_z(r,\varphi,z,k)
\end{pmatrix}
=e^{-i k z/\beta}\sum\limits_{m=0}^\infty
\begin{pmatrix}
H_{\varphi,m}(r,k)\\E_{r,m}(r,k)\\E_{z,m}(r,k)
\end{pmatrix}
\sin(m \varphi),\nonumber\\
\begin{pmatrix}
E_{\varphi}(r,\varphi,z,k)\\H_r(r,\varphi,z,k)\\H_z(r,\varphi,z,k)
\end{pmatrix}
=e^{-i k z/\beta}\sum\limits_{m=0}^\infty
\begin{pmatrix}
E_{\varphi,m}(r,k)\\H_{r,m}(r,k)\\H_{z,m}(r,k)
\end{pmatrix}\cos(m \varphi).
\end{align}
The electric displacement $\vec{D}$ and the magnetic induction $\vec{B}$ are defined using complex permittivity and permeability diagonal tensors
\begin{align}
\vec{D}= 
\begin{pmatrix}
\epsilon_r(r,k)&0&0\\
0&\epsilon_\varphi(r,k)&0\\
0&0&\epsilon_z(r,k)
\end{pmatrix}\vec{E},\quad
\vec{B}= 
\begin{pmatrix}
\mu_r(r,k)&0&0\\
0&\mu_\varphi(r,k)&0\\
0&0&\mu_z(r,k)
\end{pmatrix}\vec{H}.\nonumber
\end{align}
We do not have to assume any particular frequency dependence. In order to include conductivity and other losses in our code ECHO1D we use the following expressions (here we consider as example $r$-component):
\begin{align}
\epsilon_r(r,k)=\epsilon_{r}^{'}(r)(1+i\delta_r^{\epsilon}(r))+i\frac{\kappa_{r}(r)}{\omega(1+i \omega\tau_r(r))},\quad
\mu_r(r,k)=\mu_{r}^{'}(r)(1+i\delta_r^{\mu}(r)),\quad\omega=kc,\nonumber
\end{align}
where $\epsilon_{r}^{'}$ is the real part of the complex permettivity,  $\mu_{r}^{'}$ is the real part of the complex permeability, and the loss can be introduced with the help of dielectric loss tangent $\delta_r^{\epsilon}$, magnetic loss tangent $\delta_r^{\mu}$ or/and with AC conductivity following the Drude model~\cite{Jackson}, where $\kappa_{r}$ is the DC conductivity of the material and $\tau_r$ its relaxation time. We use similar expressions for $\varphi$- and $z$- components of the permittivity and the permeability tensors.

For each mode number $m$   we can write an independent system of equations
\begin{align}\label{EqMaxwellRound}
&\frac{m}{r} H_{z,m}+i\frac{k}{\beta}  H_{\varphi,m}=i \omega \epsilon_r E_{r,m},\nonumber\\
&-i \frac{k}{\beta} H_{r,m}-\frac{\partial}{\partial r}H_{z,m} =i \omega \epsilon_{\varphi} E_{\varphi,m},\nonumber\\
&\frac{1}{r}\frac{\partial}{\partial r}(r H_{\varphi,m})-\frac{m}{r} H_{r,m}=i \omega \epsilon_z E_{z,m}+v \rho_m,\nonumber\\
&-\frac{m}{r} E_{z,m}+i\frac{k}{\beta}  E_{\varphi,m}=-i \omega \mu_r H_{r,m},\nonumber\\
&-i \frac{k}{\beta} E_{r,m}-\frac{\partial}{\partial r}E_{z,m} =-i \omega \mu_{\varphi} H_{\varphi,m},\nonumber\\
&\frac{1}{r}\frac{\partial}{\partial r}(r E_{\varphi,m})+\frac{m}{r} E_{r,m}=-i \omega \mu_z H_{z,m},\nonumber\\
&\frac{1}{r}\frac{\partial}{\partial r}(r H_{r,m}\mu_r)-\frac{m}{r} H_{\varphi,m}\mu_{\varphi}-i k H_{z,m}\mu_z=0,\nonumber\\
&\frac{1}{r}\frac{\partial}{\partial r}(r E_{r,m}\epsilon_r)+\frac{m}{r} E_{\varphi,m}\epsilon_{\varphi}-i k E_{z,m}\epsilon_z=\rho_m.
\end{align}
We have reduced the initial three-dimensional problem to an infinite set of 
independent dimensional problems, Eqs.~(\ref{EqMaxwellRound}), for the Fourier componets of the field. 

In rectangular case we choose a coordinate system with $y$ in the vertical and $x$ in the horizontal directions; the $z$ coordinate is directed along the beam direction. The structures considered in this paper have constant width $2w$ in $x$-direction between two  perfectly conducting side walls. 

The charge density can be expanded in Fourier series
\begin{align}\label{}
\rho(x,y,z,k)=\frac{e^{-i k z/\beta}}{w}\sum\limits_{m=1}^\infty\rho_m(y)\sin(k_{x,m} x_0)\sin(k_{x,m} x),\quad k_{x,m} =\frac{\pi m}{2w},\quad
\rho_m(y)=\frac{q\delta(y-y_0)}{v},\nonumber
\end{align}
where $x_0,y_0$ are coordinates of the point charge.
Again it follows from the linearity of Maxwell's equations that the components of electromagnetic field can be represented by infinite sums:
\begin{align}\label{}
\begin{pmatrix}
H_x(x,y,z,k)\\E_y(x,y,z,k)\\E_z(x,y,z,k)
\end{pmatrix}
=\frac{e^{-i k z/\beta}}{w}\sum\limits_{m=1}^\infty
\begin{pmatrix}
H_{x,m}(y,k)\\E_{y,m}(y,k)\\E_{z,m}(y,k)
\end{pmatrix}
\sin(k_{x,m} x),\nonumber\\
\begin{pmatrix}
E_x(x,y,z,k)\\H_y(x,y,z,k)\\H_z(x,y,z,k)
\end{pmatrix}
=\frac{e^{-i k z/\beta}}{w}\sum\limits_{m=1}^\infty
\begin{pmatrix}
E_{x,m}(y,k)\\H_{y,m}(y,k)\\H_{z,m}(y,k)
\end{pmatrix}
\cos(k_{x,m} x).\nonumber
\end{align}
For each mode number $m$   we can write an independent system of equations
\begin{align}\label{EqMaxwellRect}
&-k_{x,m} H_{z,m}+i\frac{k}{\beta}  H_{x,m}=i \omega \epsilon_y E_{y,m},\nonumber\\
&-i \frac{k}{\beta} H_{y,m}-\frac{\partial}{\partial y}H_{z,m} =i \omega \epsilon_{x} E_{x,m},\nonumber\\
&\frac{\partial}{\partial y}H_{x,m}+k_{x,m} H_{y,m}=i \omega \epsilon_z E_{z,m}+v\rho_m,\nonumber\\
&k_{x,m} E_{z,m}+i\frac{k}{\beta}  E_{x,m}=-i \omega \mu_y H_{y,m},\nonumber\\
&-i \frac{k}{\beta} E_{y,m}-\frac{\partial}{\partial y}E_{z,m} =-i \omega \mu_{x} H_{x,m},\nonumber\\
&\frac{\partial}{\partial y}(E_{x,m})-k_{x,m} E_{y,m}=-i \omega \mu_z H_{z,m},\nonumber\\
&\frac{\partial}{\partial y}(H_{y,m}\mu_y)+k_{x,m} H_{x,m}\mu_{x}-i k H_{z,m}\mu_z=0,\nonumber\\
&\frac{\partial}{\partial y}(E_{y,m}\epsilon_y)-k_{x,m} E_{x,m}\epsilon_{x}-i k E_{z,m}\epsilon_z=\rho_m.
\end{align}
	
We are interested in coupling impedances as defined in~\cite{Chao93, Mounet12}. For round pipe the coupling impedance can be written as
\begin{align}\label{EqImpRoundWG}
&Z_{\parallel}(r_0,\varphi_0,r,\varphi,k,\gamma)
=\sum\limits_{m=0}^\infty Z_m(k,\gamma)I_m\left(\frac{k r_0}{\gamma\beta}\right) I_m\left(\frac{k r}{\gamma\beta}\right)\cos(m(\varphi-\varphi_0))+Z_{sc}(r_0,\varphi_0,r,\varphi,k,\gamma),\nonumber\\
&Z_{sc}(r_0,\varphi_0,r,\varphi,k,\gamma)=-\frac{k Z_0}{2\pi(\gamma^2-1)}K_0\left(\frac{k\sqrt{r_0^2+r^2-2r_0 r cos(\varphi-\varphi_0)}}{\gamma\beta}\right),
\end{align}
where $\gamma$ is the relative relativistic energy and we have written explicitly the space charge contribution $Z_{sc}$.

For a rectangular pipe the impedance reads
\begin{align}\label{EqImpRectWG}
Z_{\parallel}(x_0,y_0,x,y,k)=
\frac{1}{w}\sum\limits_{m=1}^\infty Z_m(y_0,y,k,\gamma)\sin(k_{x,m} x_0)\sin(k_{x,m} x)+Z_{sc}(x_0,y_0,x,y,k,\gamma),\nonumber\\
Z_{sc}(x_0,y_0,x,y,k,\gamma)=-\frac{k Z_0}{2\pi(\gamma^2-1)}K_0\left(\frac{k\sqrt{(x-x_0)^2+(y-y_0)^2}}{\gamma\beta}\right),
\end{align}
where
\begin{align}
Z_m(y_0,y,k,\gamma)=\left[Z_m^{cc}(k,\gamma)\cosh(k_{y,m} y_0)
+Z_m^{sc}(k,\gamma)\sinh(k_{y,m} y_0)\right]\cosh(k_{y,m} y)
\nonumber\\
+\left[Z_m^{cs}(k,\gamma)\cosh(k_{y,m} y_0)+Z_m^{ss}(k,\gamma)\sinh(k_{y,m} 
y_0)\right]\sinh(k_{y,m} y),\nonumber\\
k_{y,m}=\sqrt{k_{x,m}^2+\frac{k^2}{\gamma^2\beta^2}}.\nonumber
\end{align}

In Eqs.(\ref{EqImpRoundWG}, \ref{EqImpRectWG}) the infinite sum defines a so-called wall impedance. The longitudinal and the transverse wall impedances are connected by Panofsky-Wentzel theorem (see~\cite{Mounet12} for a detailed discussion).

The wake field effect in time domain is described by a longitudinal wake function which can be obtained by the Fourier transform of the longitudinal impedance
\begin{align}
w_{||}(s)=\frac{c}{2\pi}\int_{-\infty}^{\infty} Z_{||}(k) e^{i k s/\beta}dk,\nonumber
\end{align}
where $s$ is the distance between the source and the test particles~\cite{Chao93}.

%
\section{Field matching for uniaxial anisotropy}\label{sec:3}


In the general anisotropic case from system of first-order Eqs.(\ref{EqFieldsRound}) we obtain the second-order coupled equations for $z$-components of the electric and the magnetic fields:
\begin{align}\label{EqE2}
\frac{1}{r}\frac{\partial}{\partial r}\frac{r \epsilon_r }{\nu_{r\varphi}^2}\frac{\partial}{\partial r}E_{z,m}-\left(\frac{m^2\epsilon_{\varphi}}{r^2 \nu_{\varphi r}^2} +\epsilon_z\right)E_{z,m}+\frac{m}{r v}\left[\frac{\partial}{\partial r}\frac{1 }{\nu_{r\varphi}^2}-\frac{1 }{\nu_{\varphi r}^2}\frac{\partial}{\partial r}\right]H_{z,m}=\frac{i q\delta(r-r_0)}{\pi r_0 (1+\delta_{m0})\omega},\nonumber\\
\frac{1}{r}\frac{\partial}{\partial r}\frac{r \mu_r }{\nu_{\varphi r}^2}\frac{\partial}{\partial r}H_{z,m}-\left(\frac{m^2\mu_{\varphi}}{r^2 \nu_{r\varphi}^2} +\mu_z\right)H_{z,m}-\frac{m}{r v}\left[\frac{\partial}{\partial r}\frac{1 }{\nu_{\varphi r}^2}-\frac{1 }{\nu_{r\varphi}^2}\frac{\partial}{\partial r}\right]E_{z,m}=0,\nonumber\\
\nu_{r\varphi}^2=k^2\beta^{-2}-\omega^2\epsilon_r^2\mu_\varphi^2,\qquad \nu_{\varphi r}^2=k^2\beta^{-2}-\omega^2\epsilon_\varphi^2\mu_r^2.
\end{align}

The field matching technique for round and flat {\it isotropic} pipes was considered, for example, in~\cite{Mounet12, Piot12,Burov,Burov2,Ivanyan,Metral}. For the case of {\it uniaxial anisotropy} along $z$-axis we use the same technique, which we describe shortly in this Section.  
\begin{figure}[htbp]
	\centering
	\includegraphics*[height=60mm]{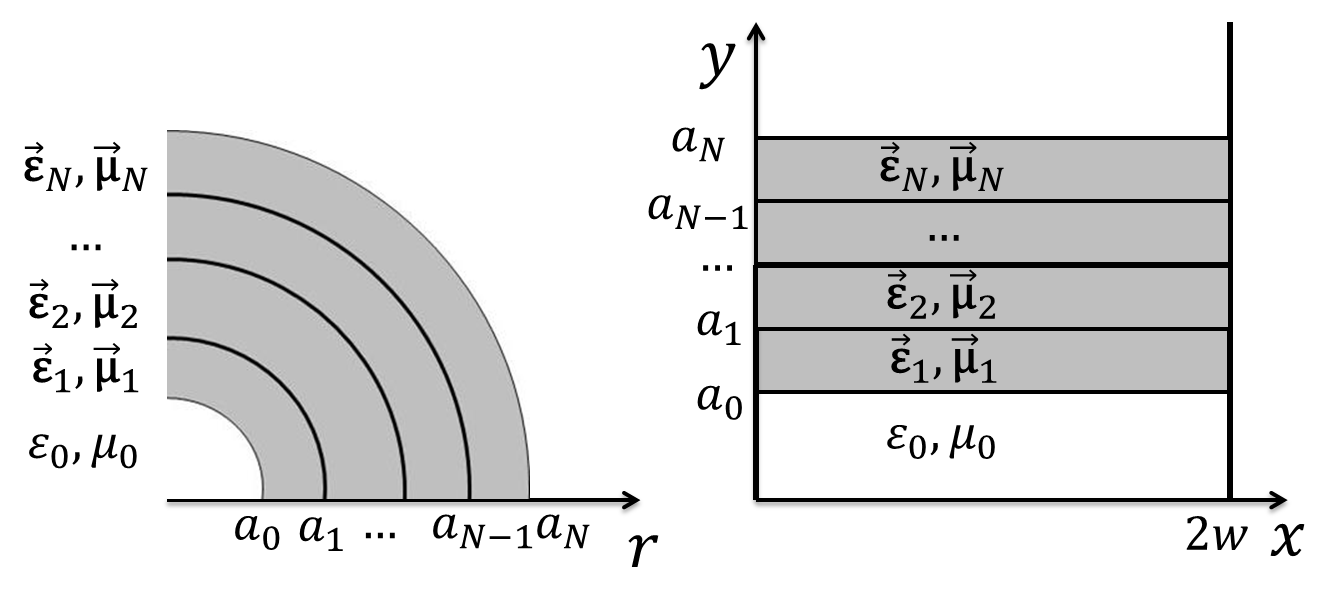}
	\caption{Examples of "round" and "rectangular" layered geometry.}\label{Fig02}
\end{figure}

We consider the uniaxial anisotropy when the permittivity and the permeability tensors are diagonal and for their elements the following relations hold
\begin{align}
\epsilon_r(r)=\epsilon_{\varphi}(r),\qquad \mu_r(r)=\mu_{\varphi}(r).\nonumber
\end{align}
Inside of each layer where the complex permeability and permittivity are constants (independent from $r$)   Eqs.(\ref{EqE2}) reduce to the decoupled equations 
\begin{align}
\label{EqE2h}
\frac{1}{r}\frac{\partial}{\partial r}r\frac{\partial}{\partial r}E_{z,m}-\left(\frac{m^2}{r^2 } +\nu_{r}^2\frac{\epsilon_z}{\epsilon_r}\right)E_{z,m}=\frac{i q\delta(r-r_0)\nu_{r}^2}{\pi r_0 (1+\delta_{m0})\omega\epsilon_r},\nonumber\\
\frac{1}{r}\frac{\partial}{\partial r}r\frac{\partial}{\partial r}H_{z,m}-\left(\frac{m^2}{r^2} +\nu_{r}^2\frac{\mu_z}{\mu_r}\right)H_{z,m}=0,\nonumber\\
\nu_{r}^2=k^2\beta^{-2}-\omega^2\epsilon_r^2\mu_r^2.
\end{align}
A general solution of homogeneous hyperbolic Eqs.~(\ref{EqE2h}) can be written in form
\begin{align}
\label{EqSolEH}
E_{z,m}(r)=C_I^m I_m(\nu_r^\epsilon r)+C_K^m K_m(\nu_r^\epsilon r),\quad
H_{z,m}(r)=D_I^m I_m(\nu_r^\mu r)+D_K^m K_m(\nu_r^\mu r),\\
\nu_r^\epsilon=\nu_r\sqrt{\epsilon_z/\epsilon_r},\quad \nu_r^\mu=\nu_r\sqrt{\mu_z/\mu_r},\nonumber
\end{align}
where $I_m, K_m$ are modified Bessel functions of complex argument.

In the following we will numerate the layers by index $j$ and $r=a_j$ defines interface between the layers with numbers$j$ and $j+1$. In order to find the constants $C_I^{m,j},C_K^{m,j},D_I^{m,j},D_K^{m,j}$ in Eqs.~(\ref{EqSolEH}) we can use 4 conditions at the interfaces between the layers:
\begin{align}
\label{EqContinuety}
E_{z,m}^j (a_j)=E_{z,m}^{j+1}(a_j),\qquad
H_{z,m}^j (a_j)=H_{z,m}^{j+1}(a_j),\nonumber\\
\epsilon_r^j E_{r,m}^j (a_j)=\epsilon_r^{j+1}E_{r,m}^{j+1} (a_j),\qquad
\mu_r^j H_{r,m}^j (a_j)=\mu_r^{j+1}H_{r,m}^{j+1} (a_j),
\end{align}
where the radial field components are defined through the longitudinal ones as
\begin{align}
\label{EqTransField}
E_{r,m}^j (r)=\frac{i k}{\nu_r^2}\left(
\frac{1}{\beta}\frac{\partial}{\partial r}E_{m,z}^j+\frac{m c \mu_r}{r}H_{z,m}
\right),\nonumber\\
H_{r,m}^j (r)=\frac{i k}{\nu_r^2}\left(
\frac{1}{\beta}\frac{\partial}{\partial r}H_{m,z}^j+\frac{m c \epsilon_r}{r}E_{z,m}
\right).
\end{align}
From Eqs.~(\ref{EqSolEH})-(\ref{EqTransField}) at each interface $r=a_j$ we obtain the relations
\begin{align}
(C_I^{m,j+1},C_K^{m,j+1},D_I^{m,j+1},D_K^{m,j+1})^T={\bf M}_j(C_I^{m,j},C_K^{m,j},D_I^{m,j},D_K^{m,j})^T,\nonumber
\end{align}
where ${\bf M}_j$ is a complex matrix of order 4. We do not write the explicit form of the elements of the matrix ${\bf M}_j$. They can be written as a combination of modified Bessel functions and the expressions are similar to those obtained in~\cite{Mounet12} for an isotropic case.

The matrix connecting the coefficients from vacuum layer to the coefficients of the last layer can be found as a matrix product
  \begin{align}
  {\bf M}={\bf M}_{N-1}{\bf M}_{N-2}...{\bf M}_{1}{\bf M}_{0}.\nonumber
  \end{align}

From the boundary condition at the axis we have $D_K^{m,0}=0$. If the last layer, $j=N$, is infinite with finite conductivity then we have open boundary condition. The field should decay at infinity, giving $C_K^{m,N}=0,D_K^{m,N}=0$, and we are looking for the solution of the following simple system
	\begin{align}
	\label{EqMopen}
	\begin{pmatrix}
	M_{11}&M_{13}&0&0\\
	M_{21}&M_{23}&-1&0\\
	M_{31}&M_{33}&0&0\\
	M_{41}&M_{43}&0&-1
	\end{pmatrix}
	\begin{pmatrix}
	C_I^{m,0}/C_K^{m,0}\\
	D_I^{m,0}/C_K^{m,0}\\
	C_K^{m,N}/C_K^{m,0}\\
	D_K^{m,N}/C_K^{m,0}
	\end{pmatrix}
	=\begin{pmatrix}
	-M_{12}\\
	-M_{22}\\
	-M_{32}\\
	-M_{42}
	\end{pmatrix}.
	\end{align}
After numerically solving of Eq.~(\ref{EqMopen}) the modal longitudinal impedance in Eq.~(\ref{EqImpRoundWG}) can be found as
\begin{align}
Z_m(k,\gamma)=-\frac{i k Z_0}{2\pi(\gamma^2-1)}\frac{C_I^{m,0}}{C_K^{m,0}}.\nonumber
\end{align}
If the last layer, $j=N$, is closed with perfectly electric conducting (PEC) material at $r=a_{N}$, then we use a modified matrix
 \begin{align}
{\bf M}={\bf M}_{N}^{C2F}{\bf M}_{N-1}{\bf M}_{N-2}...{\bf M}_{1}{\bf M}_{0},\nonumber
\end{align} 
where ${\bf M}_{N}^{C2F}$ is a matrix converting the field coefficients in the field components $H_r,H_{\varphi}$ and their derivatives:
 \begin{align}
 \begin{pmatrix}
H_{r,m}(a_{N})\\
H_{\varphi,m}(a_{N})\\
\frac{\partial}{\partial r}[H_{\varphi,m}r]|_{r=a_{N}}\\
\frac{\partial}{\partial r}[\mu_r H_{r,m}r]|_{r=a_{N}}
 \end{pmatrix}
={\bf M}_{N}^{C2F}
\begin{pmatrix}
C_I^{m,N}\\
C_K^{m,N}\\
D_I^{m,N}\\
D_K^{m,N}
\end{pmatrix}.\nonumber
 \end{align}
Again we do not write the explicit form of the elements of the matrix ${\bf M}_{N}^{C2F}$. They can be written as a combination of modified Bessel functions and the expressions are easy to obtain from Eqs.~(\ref{EqMaxwellRound}), (\ref{EqSolEH}) in any computer program supporting symbolic calculations. 
 
The boundary conditions for perfectly conducting material at $a_{N}$ can be written as $H_{r,m}(a_{N})=0$, $\frac{\partial}{\partial r}[H_{\varphi,m}r]|_{r=a_{N}}=0$.  Hence in order to find the impedance we again use Eqs.~(\ref{EqMopen}) where the right hand side has the same form but the vector of unknowns is different: $(C_I^{m,0}/C_K^{m,0},
D_I^{m,0}/C_K^{m,0},
H_{\varphi,m}(a_{N})/C_K^{m,0},\frac{\partial}{\partial r}[\mu_r H_{r,m}r]|_{r=a_{N}}/C_K^{m,0})^T$.
 
For rectangular geometries we follow the same approach. The field in the homogeneous uniaxially anisotropic layer can be presented as sum of complex exponents
 \begin{align}
 E_{z,m}(r)=C_+^m e^{k_{y,m}^\epsilon y}+C_-^m e^{-k_{y,m}^\epsilon y},\qquad
 H_{z,m}(r)=D_+^m e^{k_{y,m}^\mu y}+D_-^m e^{-k_{y,m}^\mu y},\nonumber\\
 k_{y,m}^\epsilon=\sqrt{k_{x,m}^2+\nu_y^2\frac{\epsilon_z}{\epsilon_y}},\quad  k_{y,m}^\mu=\sqrt{k_{x,m}^2+\nu_y^2\frac{\mu_z}{\mu_y}},\quad
 \nu_y^2=k^2\beta^{-2}-\omega^2\epsilon_y^2\mu_y^2.\nonumber
 \end{align}
In the following we consider only the case where the rectangular structure is symmetric in the $y$-direction (up-bottom symmetry). In this case Eq.~(\ref{EqImpRectWG}) has a simpler form
\begin{align}
Z_m(y_0,y,k,\gamma)=Z_m^{cc}(k,\gamma)\cosh(k_{y,m} y_0)
\cosh(k_{y,m} y)
+Z_m^{ss}(k,\gamma)\sinh(k_{y,m} 
y_0)\sinh(k_{y,m} y).\nonumber
\end{align}
The item $Z_m^{cc}(k,\gamma)$ can be found from the solution of the problem in the half of the domain with magnetic boundary condition at the symmetry plane $H_{z,m}(0)=0$. 
If the last layer, $j=N$, is infinite with finite conductivity then we have open boundary condition. The field should decay at infinity and it results in $C_+^{m,N}=0,D_+^{m,N}=0$. Hence we are looking for the solution of the following system
\begin{align}
\label{EqFMR1}
\begin{pmatrix}
M_{11}+M_{12}&M_{13}-M_{14}&0&0\\
M_{21}+M_{22}&M_{23}-M_{24}&-1&0\\
M_{31}+M_{32}&M_{33}-M_{34}&0&0\\
M_{41}+M_{42}&M_{43}-M_{44}&0&-1
\end{pmatrix}
\begin{pmatrix}
C_+^{m,0}/(C_-^{m,0}-C_+^{m,0})\\
D_+^{m,0}/(C_-^{m,0}-C_+^{m,0})\\
C_-^{m,N}/(C_-^{m,0}-C_+^{m,0})\\
D_-^{m,N}/(C_-^{m,0}-C_+^{m,0})
\end{pmatrix}
=\begin{pmatrix}
-M_{12}\\
-M_{22}\\
-M_{32}\\
-M_{42}
\end{pmatrix}.
\end{align}
After numerical solution of Eq.~(\ref{EqFMR1}) the item $Z_m^{cc}(k,\gamma)$ can be found as
\begin{align}
Z_m^{cc}(k,\gamma)=-\frac{2i k Z_0}{\pi(\gamma^2-1) k_{y,m}^{0}} \frac{C_+^{m,0}}{(C_-^{m,0}-C_+^{m,0})},\qquad
k_{y,m}^{0}=\sqrt{k_{x,m}^2+\frac{k^2}{\gamma^2\beta^2}}.\nonumber
\end{align}
The item $Z_m^{ss}(k,\gamma)$ can be found from the solution of another  problem in the half of the domain with electric boundary condition at the symmetry plane $E_{z,m}(0)=0$. We are looking for the solution of the following system
\begin{align}
\label{EqFMR2}
\begin{pmatrix}
M_{11}-M_{12}&M_{13}+M_{14}&0&0\\
M_{21}-M_{22}&M_{23}+M_{24}&-1&0\\
M_{31}-M_{32}&M_{33}+M_{34}&0&0\\
M_{41}-M_{42}&M_{43}+M_{44}&0&-1
\end{pmatrix}
\begin{pmatrix}
C_+^{m,0}/(C_-^{m,0}+C_+^{m,0})\\
D_+^{m,0}/(C_-^{m,0}+C_+^{m,0})\\
C_-^{m,N}/(C_-^{m,0}+C_+^{m,0})\\
D_-^{m,N}/(C_-^{m,0}+C_+^{m,0})
\end{pmatrix}
=\begin{pmatrix}
-M_{12}\\
-M_{22}\\
-M_{32}\\
-M_{42}
\end{pmatrix}.
\end{align}
After numerical solution of Eq.(\ref{EqFMR2}) the  item $Z_m^{ss}(k,\gamma)$ can be found as 
\begin{align}
Z_m^{ss}(k,\gamma)=-\frac{2i k Z_0}{\pi(\gamma^2-1) k_{y,m}^{0}} \frac{C_+^{m,0}}{(C_-^{m,0}+C_+^{m,0})}.\nonumber
\end{align}	

If the last layer, $j=N$, is closed with perfectly conducting material at $y=a_{N}$ then we use a modified matrix in the same way as described above for the round geometry. We will not consider here a rectangular structure without symmetry. In general, matrix ${\bf M}$ is a composition of matrices for all layers. It can be found and treated  in the same way as described in~\cite{Mounet12} for an isotropic case.

%
\section{Finite-difference method for full anisotropy}\label{sec:4}
%
In this section we describe a finite-difference method to treat the round and the rectangular structures with arbitrary anisotropic materials. We start with the round case. At the beginning we have to decide which equations to use. The system (\ref{EqMaxwellRound}) contains 8 first-order equations for 6 unknown field components. It can be reduced only to 2 second-order equations. For example we can use Eqs.~(\ref{EqE2}) for longitudinal components of electric and magnetic fields. However for relativistic beam in vacuum these equations degenerate: the coefficients in highest derivatives go to infinity. We would like to have a pair of equations which are non-singular and give the field components even in a perfectly conducting vacuum pipe. The relativistic charge in the limit $v=c$ in perfectly  conducting pipe does not have the longitudinal filed components. Hence the equations should be ones for the transverse field components. A possible choice could be to write equations for the radial components of electric and magnetic fields. However for higher order modes, $m>0$, these equations have singular coefficients as well.

We suggest to solve the well-posed problem for transverse components of magnetic field only,
\begin{align}
\label{EqMagnField1}
\frac{\partial}{\partial r}\frac{1}{r \epsilon_z}\frac{\partial}{\partial r}[H_{\varphi,m}^s r]+b_\varphi(r)[H_{\varphi,m}^s r]+\frac{m}{r^2\epsilon_r\mu_z}\frac{\partial}{\partial r}[\mu_r H_{r,m} r]-\nonumber\\ 
\frac{\partial}{\partial r}\left(\frac{m}{r^2\epsilon_z\mu_r}[\mu_r H_{r,m} r]\right)=-b_\varphi(r)[H_{\varphi,m}^0 r],
\end{align}
\begin{align}
\label{EqMagnField2}
\frac{\partial}{\partial r}\frac{1}{r\mu_z}\frac{\partial}{\partial r}[\mu_r H_{r,m} r]+b_r(r)[\mu_r H_{r,m}r]
+\frac{m\epsilon_{\varphi}}{r^2\epsilon_z}\frac{\partial}{\partial r}[H_{\varphi,m}^sr]-\nonumber\nonumber\\
\frac{\partial}{\partial r}\left(
\frac{m\mu_{\varphi}}{r^2\mu_z}[H_{\varphi,m}^s r]
\right)=\frac{\partial}{\partial r}\left(\frac{m\mu_{\varphi}}{r\mu_z}H_{\varphi,m}^0\right),
\end{align}
\begin{align}
b_\varphi(r)=\frac{\omega^2\mu_\varphi}{r}-\frac{k^2}{r\epsilon_r\beta^2}-\frac{m^2\mu_\varphi}{r^3\epsilon_r\mu_z},\quad
b_r(r)=\frac{\omega^2\epsilon_\varphi}{r}-\frac{k^2}{r\mu_r\beta^2}-\frac{m^2\epsilon_\varphi}{r^3\mu_r\epsilon_z}.\nonumber
\end{align}
In order to remove the discontinuity of the azimuthal component in the charge location $r_0$ we present the azimuthal component of the magnetic field in the form
\begin{align}
  H_{\varphi,m}=H_{\varphi,m}^s+H_{\varphi,m}^0,\quad H_{\varphi,m}^0=(1+\delta_{m0})H_{\varphi}^0,\quad H_{\varphi}^0=\frac{\theta(r-r_0)}{2\pi r},\nonumber
\end{align}
where $\theta(r)$ is Heaviside function and $H_{\varphi}^0$ presents a monopole harmonic of  the self field of relativistic charge in free space. Let us note that $H_{\varphi,m}^s$ has the meaning of the scattered field only for the lowest monopole mode, $m=0$, and the relativistic charge. Another choice could be to take $H_{\varphi,m}^0$ as a true $m$-harmonic of the self-field but this introduces additional terms into the right-hand side of Eqs.~(\ref{EqMagnField1}), (\ref{EqMagnField2}) without any clear improvement of the accuracy of the numerical solution.

\begin{figure}[htbp]
	\centering
	\includegraphics*[height=40mm]{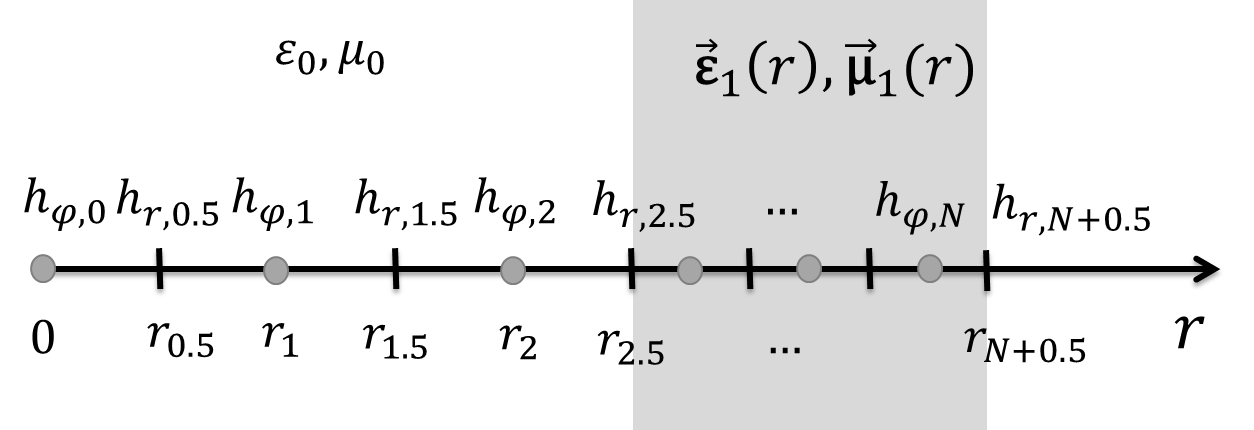}
	\caption{One dimensional mesh and positions of the transverse magnetic field components.}\label{FigMesh}
\end{figure}

\begin{figure}[htbp]
	\centering
	\includegraphics*[height=40mm]{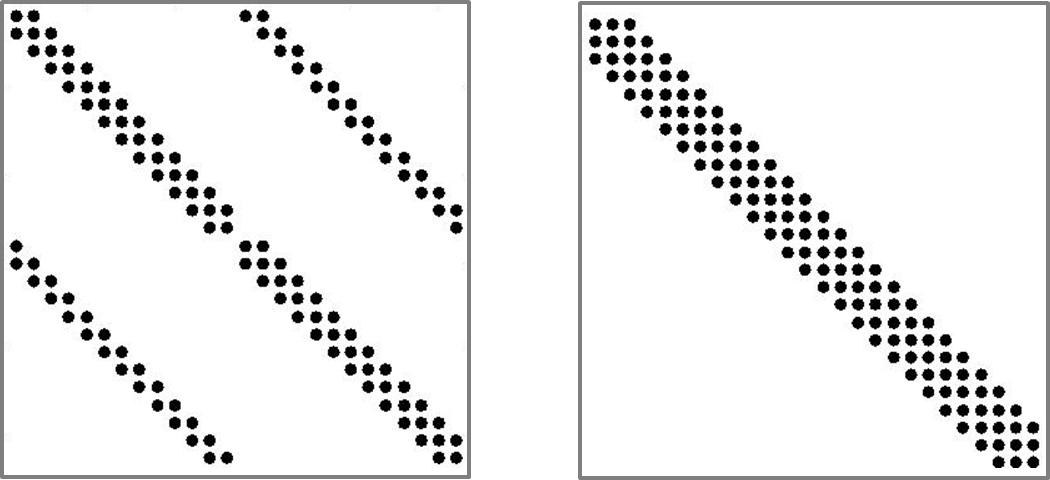}
	\caption{Reduction of seven band matrix to pentadiagonal form.}\label{FigBand}
\end{figure}

We introduce one dimensional mesh with shifted positions of the transverse magnetic filed components as shown in Fig.~\ref{FigMesh}. The mesh in material is not equidistant in general. It is chosen to sample the wave length in the material properly and depends on the wavenumber $k=\omega/c$. We use the standard second order approximations of the derivatives~\cite{Samarski} and the finite-difference scheme reads
\begin{align}
\label{EqFDscheme}
\frac{1}{r_{i+0.5}-r_{i-0.5}}\left[a_\varphi(r_{i+0.5})\frac{h_{\varphi,i+1}-h_{\varphi,i}}{r_{i+1}-r_{i}}-a_\varphi(r_{i-0.5})\frac{h_{\varphi,i}-h_{\varphi,i-1}}{r_{i}-r_{i-1}}\right]+b_\varphi(r_i)h_{\varphi,i}+\nonumber\\
c_\varphi(r_i)\frac{h_{r,i+0.5}-h_{r,i-0.5}}{r_{i+0.5}-r_{i-0.5}}-\frac{d_\varphi(r_{i+0.5})h_{r,i+0.5}-d_\varphi(r_{i-0.5})h_{r,i-0.5}}{r_{i+0.5}-r_{i-0.5}}=f_\varphi(r_i),\nonumber\\
\frac{1}{r_{i}-r_{i-1}}\left[a_r(r_{i})\frac{h_{r,i+0.5}-h_{r,i-0.5}}{r_{i+0.5}-r_{i-0.5}}
-a_r(r_{i-1})\frac{
h_{r,i-0.5}-h_{r,i-1.5}}{r_{i-0.5}-r_{i-1.5}}\right]
+b_r(r_{i-0.5})h_{r,i-0.5}+\nonumber\\
c_r(r_{i-0.5})\frac{h_{\varphi,i}-h_{\varphi,i-1}}{r_{i}-r_{i-1}}-
\frac{d_r(r_{i})h_{\varphi,i}-d_r(r_{i-1})h_{\varphi,i-1}}{r_{i}-r_{i-1}}=f_r(r_{i-0.5}),
\end{align}
where we have introduced the discrete field components $h_{\varphi,i}=H_{\varphi,m}^s(r_i)r_i, h_{r,i+0.5}=\mu_r(r_{i+0.5})H_{r,m}(r_{i+0.5})r_{i+0.5} $ and the following notation
\begin{align}
a_\varphi(r_{i+0.5})=\frac{1}{r_{i+0.5} \epsilon_z(r_{i+0.5})},\quad
c_\varphi(r_{i})=\frac{m}{r_i^2\epsilon_r(r_i)\mu_z(r_i)},\nonumber\\
d_\varphi(r_{i+0.5})=\frac{m}{r_{i+0.5}^2\epsilon_z(r_{i+0.5})\mu_r(r_{i+0.5})},\quad
f_\varphi(r_{i})=-b_r(r_i)[H_{\varphi,m}^0(r_i)r_i]\nonumber\\
a_r(r_{i})=\frac{1}{r_{i} \mu_z(r_{i})},\quad
c_r(r_{i-0.5})=\frac{m}{r_{i-0.5}^2\epsilon_\varphi(r_{i-0.5})\epsilon_z(r_{i-0.5})},\nonumber\\
d_r(r_{i})=\frac{m}{r_{i}^2\mu_\varphi(r_{i})\mu_z(r_{i})},\quad
f_r(r_{i-0.5})=\frac{d_r(r_{i})[H_{\varphi,m}^0(r_{i})r_{i}]-d_r(r_{i-1})[H_{\varphi,m}^0(r_{i-1})r_{i-1}]}{r_{i}-r_{i-1}}.\nonumber
\end{align}
At the axis of axially symmetric geometry we have magnetic boundary condition
\begin{align}
[H_{\varphi,r} r]|_{r=0}=0,\quad \frac{\partial}{\partial r}[\mu_r H_{r,m} r]|_{r=0}=0,\nonumber
\end{align}
and the equations for $i=1$ can be written in the form
\begin{align}
\frac{1}{r_{1.5}-r_{0.5}}\left[a_\varphi(r_{1.5})\frac{h_{\varphi,2}-h_{\varphi,1}}{r_{2}-r_{1}}-a_\varphi(r_{0.5})\frac{h_{\varphi,1}}{r_{1}}\right]+b_\varphi(r_1)h_{\varphi,1}+\nonumber\\
c_\varphi(r_1)\frac{h_{r,1.5}-h_{r,0.5}}{r_{1.5}-r_{0.5}}-\frac{d_\varphi(r_{1.5})h_{r,1.5}-d_\varphi(r_{0.5})h_{r,0.5}}{r_{1.5}-r_{0.5}}=f_\varphi(r_1),\nonumber\\
\frac{1}{r_{1}}\left[a_r(r_{1})\frac{h_{r,1.5}-h_{r,0.5}}{r_{1.5}-r_{0.5}}
\right]
+b_r(r_{0.5})h_{r,0.5}+
c_r(r_{0.5})\frac{h_{\varphi,1}}{r_{1}}-
\frac{d_r(r_{1})h_{\varphi,1}}{r_{1}}=f_r(r_{0.5}),\nonumber
\end{align}
If the exterior boundary is perfectly conducting at $r_{N+0.5}=b$ then
we have electric boundary condition for the magnetic field
\begin{align}
[\mu_r H_{r,m} r]|_{r=b}=0,\quad \frac{\partial}{\partial r}[H_{\varphi,m} r]|_{r=b}=0,\nonumber
\end{align}
and the equations for $i=N$ can be written in form~(\ref{EqFDscheme}) with $h_{\varphi,N+1}=h_{\varphi,N},h_{r,N+0.5}=0$.
Hence we have to solve a linear system
\begin{align}
\label{EqLS}
{\bf M}\vec{h}=\vec{f},\quad \vec{h}=(h_{\varphi,1},h_{\varphi,2},...,h_{\varphi,N},h_{r,0.5},h_{r,1.5},...,h_{r,N+0.5})^t,
\end{align}
where the matrix ${\bf M}$ has dimensions $2N \times 2N$ and the seven band structure shown in Fig.~\ref{FigBand} on the left side.
			
In order to use a direct method of solution of linear system (\ref{EqLS}) we introduce the permutation matrix ${\bf P}_\sigma$ defined by permutation of indexis
\begin{align}
\label{EqPermut}
\sigma_i= \left\{
\begin{array}{ll}
2N+1-\frac{i}{2} &, \quad i\quad \text{even}, \\
N+1-\frac{i-1}{2} &,\quad i\quad \text{odd}.
\end{array}
\right.
\end{align}
It converts the sparse seven band matrix  ${\bf M}$ in pentadiagonal form ${\bf P}_\sigma {\bf M} {\bf P}_\sigma^T$ shown in Fig.~\ref{FigBand} on the right side. The new system allows for a direct solution with complexity $O(N)$~\cite{Davis}, meaning that the solution time is proportional to the number of mesh points.

If the last layer is infinite then we need an open boundary condition to truncate the matrix at $r=b$. If the last material has only uniaxial anisotropy like one considered in Section~\ref{sec:3} then we can easily write such a condition. Indeed from the definition of the modified Bessel functions of the second type, the open boundary condition in the round geometry reads (see Eq.~(\ref{EqSolEH}))
\begin{align}
\label{EqOBCz}
\frac{\partial}{\partial r} E_{z,m}+\frac{K_m^\epsilon}{r}E_{z,m}=0,\quad
K_m^\epsilon=\left(m+r\nu_r^\epsilon\frac{ K_{m-1}(\nu_r^\epsilon r)}{K_{m}(\nu_r^\epsilon r)}\right),\nonumber\\
\frac{\partial}{\partial r} H_{z,m}+\frac{K_m^\mu}{r}H_{z,m}=0,\quad
K_m^\mu=\left(m+r\nu_r^\mu\frac{ K_{m-1}(\nu_r^\mu r)}{K_{m}(\nu_r^\mu r)}\right).
\end{align}
Combining Eqs.(\ref{EqOBCz}) with Maxwells equations (\ref{EqMaxwellRound}) we can derive the open boundary conditions for the transverse components of the magnetic field:
\begin{align}
\frac{\epsilon_r}{\epsilon_z}\left(K_m^\epsilon - \frac{m^2}{K_m^\mu}\right)\frac{\partial}{\partial r}[H_{\varphi,m}r]+r\nu_r^2[H_{\varphi,m}r]+
\frac{m}{r}\left(r^2\frac{\nu_r^2}{K_m^\mu}-\frac{\epsilon_r}{\epsilon_z}\left(K_m^\epsilon - \frac{m^2}{K_m^\mu}\right)\right)[\mu_r H_{r,m} r]=0,\nonumber\\
\frac{\mu_r}{\mu_z}\left(K_m^\mu - \frac{m^2}{K_m^\epsilon}\right)\frac{\partial}{\partial r}[\mu_r H_{r,m}r]+r\nu_r^2[\mu_r H_{r,m}r]+
\frac{m}{r}\left(r^2\frac{\nu_r^2}{K_m^\epsilon}-\frac{\mu_r}{\mu_z}\left(K_m^\mu - \frac{m^2}{K_m^\epsilon}\right)\right)[H_{\varphi,m} r]=0.\nonumber
\end{align}
We approximate these boundary condition on the one dimensional mesh with second order by finite differences~\cite{Samarski}. The final matrix will have the same structure as in previous situation with the perfectly conducting boundary (see Fig.\ref{FigBand}).

After numerical solution of the linear system (\ref{EqLS}) the longitudinal electric field component and the impedance can be found as
\begin{align}
E_{z,m}(r_0)=-\frac{i}{\omega\epsilon_z(r_0) r_0}\left[\frac{\partial}{\partial r}[H_{\varphi,m} r]|_{r=r_0}-m H_r(r_0)\right],\nonumber\\
Z_m(k)=\frac{E_{z,m}(r_0)}{I_m(\nu_r^0 r_0)^2}-\frac{G K_m(\nu_r^0 r_0)}{I_m(\nu_r^0 r_0)},\quad
G=\frac{i k Z_0}{2\pi(\gamma^2-1)}.\nonumber
\end{align}

In the case of rectangular geometry we again consider only the case with symmetry plane at $y=0$. In this case we have to solve two problems in half of the computational domain. The first problem for $Z_{cc}$ has a magnetic boudary condition at the symmetry plane ($H_{z,m}(0)=0$) and we approximate it in the same way as it was done at the axis for round geometry. The second problem for $Z_{ss}$ has an electric boundary condition ($E_{z,m}=0$) at the symmetry plane and we approximate it in the same way as it was done for round geometry at PEC boundary.

If the last layer of the rectangular geometry is infinite and has only uniaxial anisotropy then the
open boundary condition for the longitudinal field components read 
\begin{align}
\frac{\partial}{\partial y} E_{z,m}+k_{y,m}^\epsilon E_{z,m}=0,\quad
\frac{\partial}{\partial y} H_{z,m}+k_{y,m}^\mu H_{z,m}=0,\nonumber
\end{align}
Combining them with Maxwell's equations (\ref{EqMaxwellRect}) we can derive the open boundary conditions for the transverse components of the magnetic field in the rectangular case:
\begin{align}
\frac{\epsilon_y}{\epsilon_z}\left(k_{y,m}^\epsilon - \frac{k_{x,m}^2}{ k_{y,m}^\mu}\right)\frac{\partial}{\partial y} H_x+\nu_y^2 H_x-k_{x,m}\left(\frac{\nu_y^2}{k_{y,m}^\mu}-\frac{\epsilon_y}{\epsilon_z}\left(k_{y,m}^\epsilon - \frac{k_{x,m}^2}{k_{y,m}^\mu}\right)\right) H_{y,m}=0,\nonumber\\
\frac{\mu_y}{\mu_z}\left(k_{y,m}^\mu - \frac{k_{x,m}^2}{ k_{y,m}^\epsilon}\right)\frac{\partial}{\partial y} H_y+\nu_y^2 H_y-k_{x,m}\left(\frac{\nu_y^2}{k_{y,m}^\epsilon}-\frac{\mu_y}{\mu_z}\left(k_{y,m}^\mu - \frac{k_{x,m}^2}{k_{y,m}^\epsilon}\right)\right) H_{x,m}=0.\nonumber
\end{align}
The longitudinal electric field component and the impedance in the rectangular case  can be found as
\begin{align}
E_{z,m}(y_0)=-\frac{i}{\omega\epsilon_z(y_0)}\left[\frac{\partial}{\partial y}H_{x,m}|_{y=y_0}+k_{x,m} H_{y,m}(y_0)\right],\nonumber\\
Z_m^{cc}(k)=\frac{E_{z,m}(y_0)}{\cosh(k_{y,m}^0 y_0)^2}-\frac{2G}{k_{y,m}^0\cosh(k_{y,m}^0 y_0)}e^{-k_{y,m}^0},\nonumber\\
Z_m^{ss}(k)=\frac{E_{z,m}(y_0)}{\sinh(k_{y,m}^0 y_0)^2}-\frac{2G}{k_{y,m}^0\sinh(k_{y,m}^0 y_0)}e^{-k_{y,m}^0},\nonumber
\end{align}
where $H_{x,m},H_{y,m}$ are solutions of the corresponding problem with magnetic or electric boundary condition at the symmetry plane.

%
\section{Combination of field matching and finite-difference methods for anisotropic waveguides}\label{sec:5}
%

The finite-difference method of the previous section allows treating the full anisotropy but it could also be time-consuming as it requires a mesh in the whole domain. In this Section we suggest a combination of the field matching technique and of the finite-difference method.

\begin{figure}[htbp]
	\centering
	\includegraphics*[height=60mm]{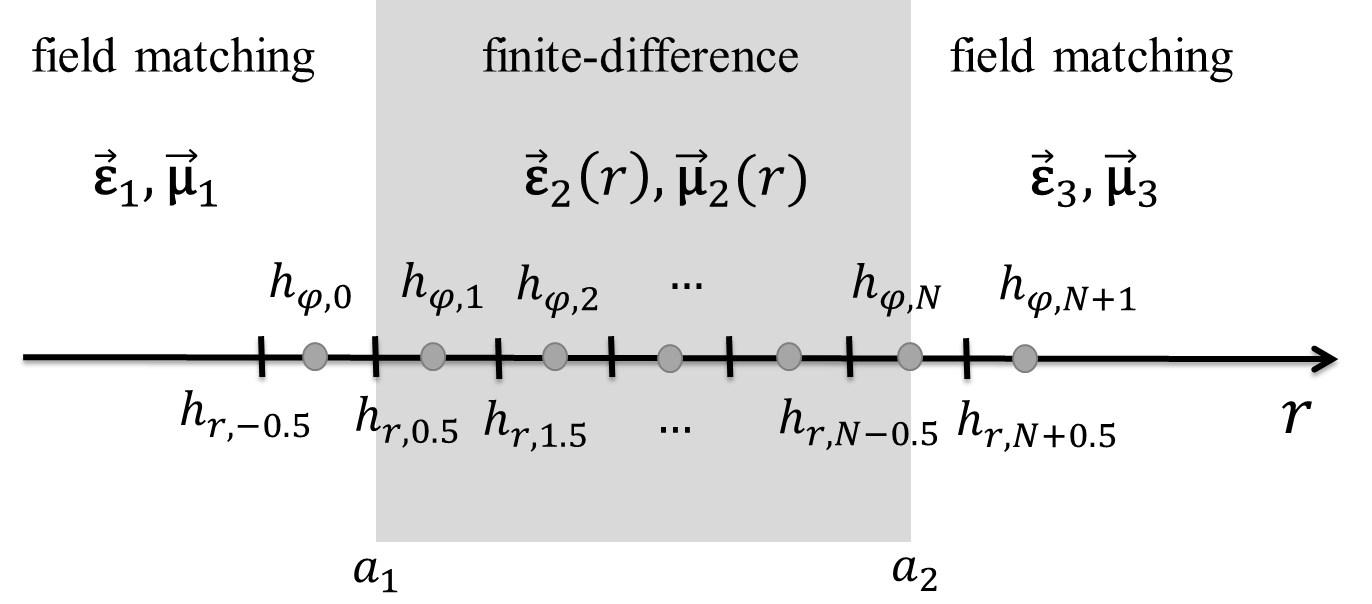}
	\caption{One dimensional mesh of combined method and positions of the transverse magnetic field components.}\label{FigFMFDmesh}
\end{figure}

Again we will start with a round geometry. In order to describe the method, let us consider example shown in Fig.\ref{FigFMFDmesh}: the first and the third layers allow solutions in analytical form, Eq.(\ref{EqSolEH}), the middle layer is anisotropic and could be treated only with finite-difference method. Let us denote the coefficients in the first layer as $C_I^{m,1},C_K^{m,1},D_I^{m,1},D_K^{m,1}$ and the coefficients in the third layer as $C_I^{m,3},C_K^{m,3},D_I^{m,3},D_K^{m,3}$. In order to use  the matrix approach of Section~\ref{sec:3} we need to find matrix ${\bf M}_{13}^{FD}$, converting the first set of coefficients in the second one:
\begin{align}
\begin{pmatrix}
C_I^{m,3}\\
C_K^{m,3}\\
D_I^{m,3}\\
D_K^{m,3}
\end{pmatrix}
={\bf M}_{13}^{FD}
\begin{pmatrix}
C_I^{m,1}\\
C_K^{m,1}\\
D_I^{m,1}\\
D_K^{m,1}
\end{pmatrix}.\nonumber
\end{align}
The matrix  ${\bf M}_{13}$ can be found as a product of several simple complex matrices of size $4\times 4$:
\begin{align}
{\bf M}_{13}={\bf M}_2^{F2C} {\bf M}_2^{F2F} {\bf M}_{12}^{FD} {\bf M}_1^{F2F} {\bf M}_1^{C2F},
\end{align}
where ${\bf M}_1^{CF}$ is the matrix introduced already at Section \ref{sec:3}. It converts the coefficients $C_I^{m,1},C_K^{m,1},D_I^{m,1},D_K^{m,1}$ in the magnetic field components (and their derivatives) $H_{r,m}(a_{1}^-),
H_{\varphi,m}(a_{1}^-),
\frac{\partial}{\partial r}[H_{\varphi,m}r]|_{r=a_{1}^-},
\frac{\partial}{\partial r}[\mu_r H_{r,m}r]|_{r=a_{1}^-}$. Here the notation $r=a_1^-$ means a one-sided limit from below. The matrices ${\bf M}_1^{F2F}$ converts the  one-sided limits of the fields components from below $H_{r,m}(a_{1}^-),
H_{\varphi,m}(a_{1}^-),
\frac{\partial}{\partial r}[H_{\varphi,m}r]|_{r=a_{1}^-},
\frac{\partial}{\partial r}[\mu_r H_{r,m}r]|_{r=a_{1}^-}$ into one-sided limits of the fields components from above $H_{r,m}(a_{1}^+),
H_{\varphi,m}(a_{1}^+),
\frac{\partial}{\partial r}[H_{\varphi,m}r]|_{r=a_{1}^+},
\frac{\partial}{\partial r}[\mu_r H_{r,m}r]|_{r=a_{1}^+}$. The matrix ${\bf M}_2^{F2F}$ makes the same at $r=a_2$. Finally matrix ${\bf M}_2^{F2C}$ converts the field components $H_{r,m}(a_{2}^+),
H_{\varphi,m}(a_{2}^+),
\frac{\partial}{\partial r}[H_{\varphi,m}r]|_{r=a_{2}^+},
\frac{\partial}{\partial r}[\mu_r H_{r,m}r]|_{r=a_{2}^+}$ into the coefficients $C_I^{m,3},C_K^{m,3},D_I^{m,3},D_K^{m,3}$. All these matrices can be found easily in the analytical form with a help of any computer program for symbolic calculations. Only the matrix ${\bf M}_{12}^{FD}$ converting  $H_{r,m}(a_{1}^+),
H_{\varphi,m}(a_{1}^+),
\frac{\partial}{\partial r}[H_{\varphi,m}r]|_{r=a_{1}^+},
\frac{\partial}{\partial r}[\mu_r H_{r,m}r]|_{r=a_{1}^+}$ into  $H_{r,m}(a_{2}^-),
H_{\varphi,m}(a_{2}^-),
\frac{\partial}{\partial r}[H_{\varphi,m}r]|_{r=a_{2}^-},
\frac{\partial}{\partial r}[\mu_r H_{r,m}r]|_{r=a_{2}^-}$ requires application of the finite-difference scheme of Section~\ref{sec:4}. 

For the combined method we use the one-dimensional mesh shown in Fig.~\ref{FigFMFDmesh}. In order to obtain the second-order approximation of the boundary conditions we use the fictive nodes outside of the layer. The equations are discretized in the same way as in Section~\ref{sec:4} for $i=1,...,N$ (see. Eq.~(\ref{EqFDscheme})) and we can write the undetermined matrix equation
\begin{align}
\label{EqLSext}
{\bf M}\vec{h}=\vec{f},\quad \vec{h}=(h_{\varphi,0},h_{\varphi,1},...,h_{\varphi,N+1},h_{r,-0.5},h_{r,0.5},...,h_{r,N+0.5})^t,
\end{align}
where ${\bf M}$ is a non-square matrix of size $2N\times(2N+4)$. In order to reduce the number of the unknowns to $2N$ we will use the boundary conditions at $r=a_1$ and exclude $h_{\varphi,0},h_{\varphi,1},h_{r,-0.5},h_{r,0.5}$. 

\begin{figure}[htbp]
	\centering
	\includegraphics*[height=40mm]{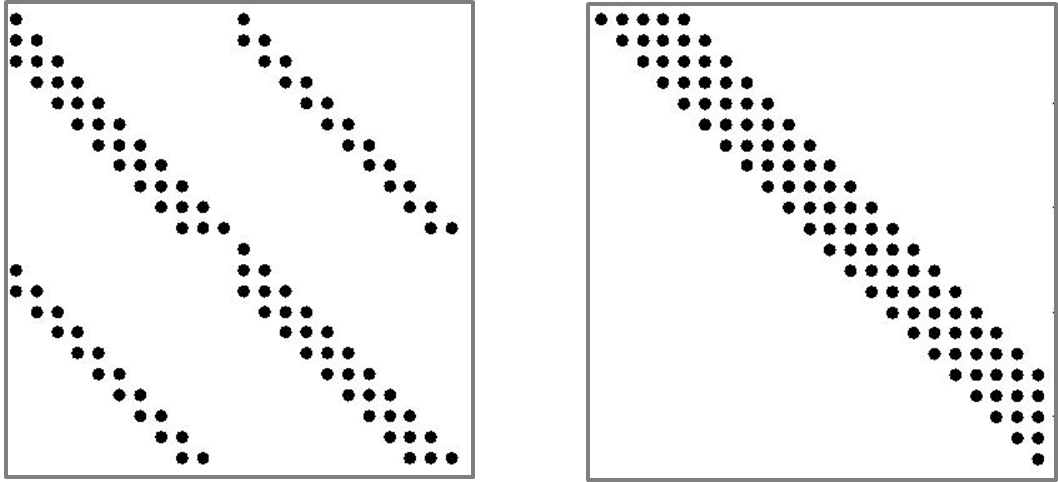}
	\caption{Reduction of seven band matrix of combined method to upper triangular form.}\label{FigBandComb}
\end{figure}

Let us write a general form of the boundary conditions at $r=a_1$
\begin{align}
\label{EqBC}
H_{r,m}(a_{1}^+)=B_r,
H_{\varphi,m}(a_{1}^+)=B_\varphi,
\frac{\partial}{\partial r}[H_{\varphi,m}r]|_{r=a_{1}^+}=D_\varphi,
\frac{\partial}{\partial r}[\mu_r H_{r,m}r]|_{r=a_{1}^+}=D_r.
\end{align}
It is easy to write the second order approximation of the first three equations~(\ref{EqBC}) and obtain the expressions for
$h_{\varphi,0},h_{\varphi,1},h_{r,0.5}$ :
\begin{align}
h_{r,0.5}=B_r,\quad
h_{\varphi,0}=B_\varphi-(r_{0.5}-r_{0})D_\varphi,\quad
h_{\varphi,1}=B_\varphi+(r_1-r_{0.5})D_\varphi.\nonumber
\end{align}
In order to find $h_{r,-0.5}$ we use the second order approximation of the fourth boundary condition and Eq.~(\ref{EqBC}) for $i=1$. After a simple algebra we obtain:
\begin{align}
h_{r,-0.5}=-\frac{M_{N+1,1}h_{\varphi,1}+M_{N+1,2}h_{\varphi,2}+M_{N+1,N+4} h_{r,0.5}+M_{N+1,N+5}D_r(r_{1.5}-r_{-0.5})}{M_{N+1,N+3}+M_{N+1,N+5}},\nonumber
\end{align}
where $M_{i,j}$ are elements of matrix ${\bf M}$ in Eq.~(\ref{EqLSext}). Through excluding of $h_{\varphi,0},h_{\varphi,1},h_{r,-0.5},h_{r,0.5}$ from system Eq.(\ref{EqLSext}) we obtain a matrix equation with reduced matrix ${\bf M}^r$ of size $2N\times 2N$:
\begin{align}
\label{EqRedM}
{\bf M}^r \vec{h}^r=\vec{f}^r,\\ \vec{h}^r=(h_{\varphi,2},h_{\varphi,3},...,h_{\varphi,N+1},h_{r,1.5},h_{r,2.5},...,h_{r,N+0.5})^t,\nonumber\\
\vec{f}^r=(f_{1}^r,f_{2}^r,f_{3},f_{N},f_{0.5}^r,f_{1.5}^r,f_{2.5}^r...,f_{N-0.5})^t,\nonumber\\
M_{i,j}^r=M_{i,j+2}, \quad i=1,...,2N, \quad j=1,...,N,\nonumber\\
M_{i,j}^r=M_{i,j+4}, \quad i=1,...,2N, \quad j=N+1,...,2N,\nonumber
\end{align}
where
\begin{align}
f_1^r=f_1-(M_{1,1}h_{\varphi,0}+M_{1,2}h_{\varphi,1}+M_{1,N+4}h_{r,0.5}),\quad
f_2^r=f_2-M_{2,2}h_{\varphi,1},\nonumber\\
f_{0.5}^r=f_{0.5}-(M_{N+1,1}h_{\varphi,0}+M_{N+1,2}h_{\varphi,1}+M_{N+1,N+3}h_{r,-0.5}+M_{N+1,N+4}h_{r,0.5}),\nonumber\\
f_{1.5}^r=f_{1.5}-(M_{N+2,2}h_{\varphi,1}+M_{N+2,N+4}h_{r,0.5}).\nonumber
\end{align}
The matrix ${\bf M}^r$ of system~(\ref{EqRedM}) has the form shown in Fig.~\ref{FigBandComb} and it can be reduced with the permutations~(\ref{EqPermut}) to the upper triangular matrix shown on the right. Hence the system requires only $O(N)$ operations to solve it.

In order to find matrix ${\bf M}_{12}^{FD}$ we need to solve the same equations but with 4 different sets of boundary conditions at $r=a_1$. The boundary conditions at $r=a_1$ for the first problem read
\begin{align}
H_{r,m}(a_{1}^+)=1,
H_{\varphi,m}(a_{1}^+)=0,
\frac{\partial}{\partial r}[H_{\varphi,m}r]|_{r=a_{1}^+}=0,
\frac{\partial}{\partial r}[\mu_r H_{r,m}r]|_{r=a_{1}^+}=0.\nonumber
\end{align}
The field components
\begin{align}
H_{r,m}(a_{2}^-)=(h_{r,N+0.5}+h_{r,N-0.5})/2,\quad
H_{\varphi,m}(a_{2}^-)=h_{\varphi,N},\nonumber\\
\frac{\partial}{\partial r}[H_{\varphi,m}r]|_{r=a_{2}^-}=\frac{h_{r,N+1}-h_{r,N-1}}{r_{N+1}-r_{N-1}},\quad
\frac{\partial}{\partial r}[\mu_r H_{r,m}r]|_{r=a_{2}^-}=\frac{h_{r,N+0.5}-h_{r,N-0.5}}{r_{N+0.5}-r_{N-0.5}}\nonumber
\end{align}
will give the elements of the first column of matrix   ${\bf M}_{12}^{FD}$.
The second column can be found from the solution of the same equations but with  another boundary condition at $r=a_1$:
\begin{align}
H_{r,m}(a_{1}^+)=0,
H_{\varphi,m}(a_{1}^+)=1,
\frac{\partial}{\partial r}[H_{\varphi,m}r]|_{r=a_{1}^+}=0,
\frac{\partial}{\partial r}[\mu_r H_{r,m}r]|_{r=a_{1}^+}=0.\nonumber
\end{align}
Analogously we will find the third and the fourth columns of this matrix.

As can be seen from the above description we need to solve the problem 4 times in the anisotropic layer only. If the layer is thin then the suggested method is faster than the finite-difference method of the previous section where the whole domain has to be discretized to sample the electromagnetic field everywhere. At the rectangular geometry the algorithm is exactly the same with corresponding equations for the rectangular case.

%
\section{Numerical examples}\label{sec:6}
%

\begin{figure}[htbp]
	\centering
	\includegraphics*[height=60mm]{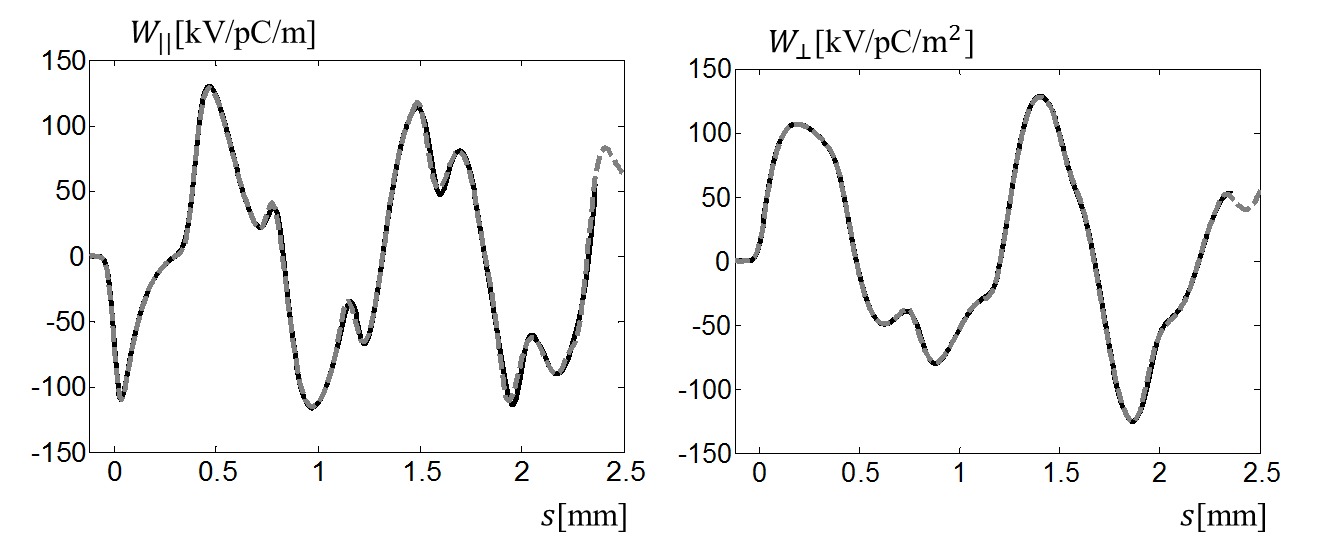}
	\caption{The longitudinal and the transverse wake potentials near the pipe axis as obtained by time-domain code ECHO2D (solid black line) and by frequency-domain code ECHO1D (grey dashed line).}\label{FigNE1}
\end{figure}

Recently, experimental demonstration of energy modulations in dielectric pipes was observed at the PITZ facility~\cite{Piot17}. The experiment was performed with a dielectric pipe with an isotropic dielectric layer of  permittivity $\epsilon=4.41\epsilon_0$. The layer starts at radius  $a_0=0.45$ mm and is closed with PEC at $a_1=0.55$ mm.
We take this dielectric pipe as our first example and calculate the steady-state wake of a relativistic Gaussian bunch with rms length $\sigma_z=25$ $\mu$m. In Fig.~\ref{FigNE1} we show the longitudinal and the transverse wake potentials near the pipe axis. The longitudinal wake potential for the charge distribution $\lambda(s)$ is defined as
\begin{align}
W_{\parallel}(s)=\int_{-\infty}^{s}w_{\parallel}(s')\lambda(s-s')ds'.\nonumber
\end{align}
The transverse wake potential is defined analogously and $W_{\perp}(s)$ means here the dipole component of the transverse wake normalized by offset~\cite{Chao93}.

\begin{figure}[htbp]
	\centering
	\includegraphics*[height=60mm]{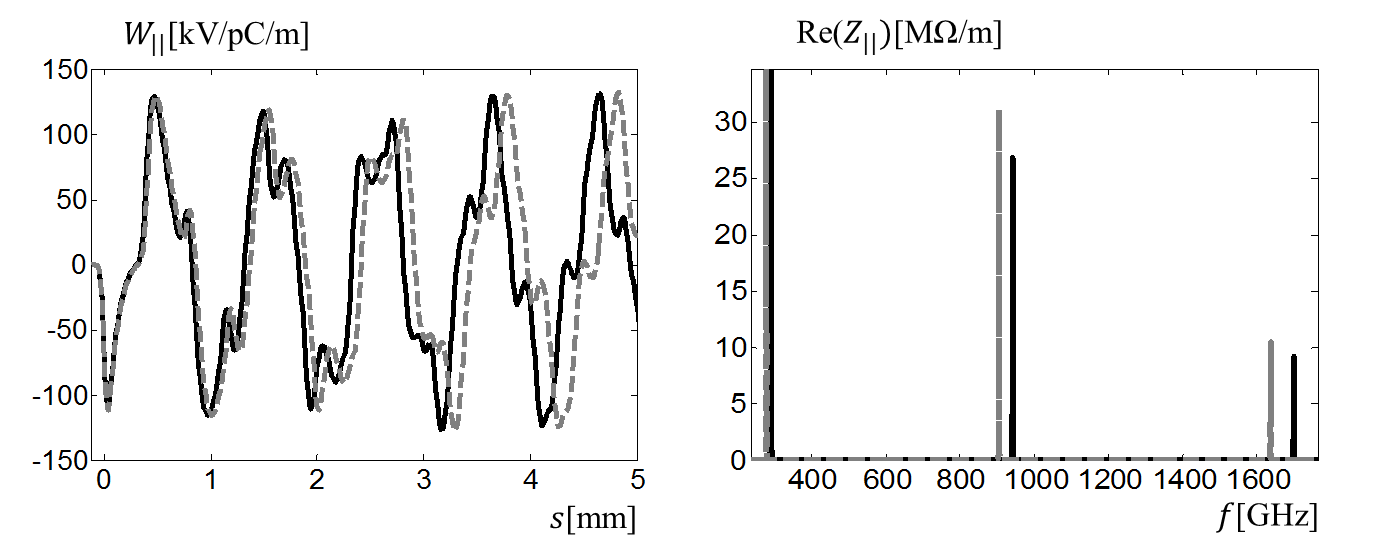}
	\caption{The longitudinal wake and the real part of the longitudinal impedance for dielectric pipe at PITZ. The solid black lines show the results for isotropic case and the dashed grey line presents the result for anysotropic case.}\label{FigNE2}
\end{figure}

The gray dashed line shows the results obtained with field matching method as described in Section~\ref{sec:3}. The solid line is obtained with time-domain code ECHO2D~\cite{Zag16}. In order to obtain the steady-state wake in time-domain we have subtracted the wake for pipe of length 10 cm from the wake of pipe of length 11 cm. The agreement of the curves from two different methods confirms the correctness of the results.
In Fig.~\ref{FigNE2} the longitudinal wake potential and the real part of the longitudinal impedance are shown. The solid black lines show the results for the isotropic case and the dashed grey line presents the result for the anisotropic case when we have changed only the permittivity in radial direction, $\epsilon_r=6\epsilon_0$. We see a clear shift in the modal frequencies for the anisotropic case. It cannot be treated with the field matching only. Here we have used methods described in Sections~\ref{sec:4},~\ref{sec:5}. The wave number $k$ was sampled from $1$ $m^{-1}$ to $10^5$ $m^{-1}$ with step 0.2. The execution times for all methods are shown in Table~\ref{Table01}. Let us note that in this example we have used a small conductivity $\kappa=1$ S/m to resolve the real part of the impedance.

\begin{table}[htbp]
	\centering
	\caption{Execution time in seconds for different methods.}
	\label{Table01}
	\begin{tabular}{lcc}
		{\bf Method}		& {\bf Round} & {\bf Rectangular} \\
		 Field Matching (Section~\ref{sec:3}) 	& 31 & 5 \\
		 Finite-Difference (Section~\ref{sec:4})	& 170& 110 \\
		 Combined (Section~\ref{sec:5})& 86 & 60\\
	\end{tabular}
\end{table}

\begin{figure}[htbp]
	\centering
	\includegraphics*[height=60mm]{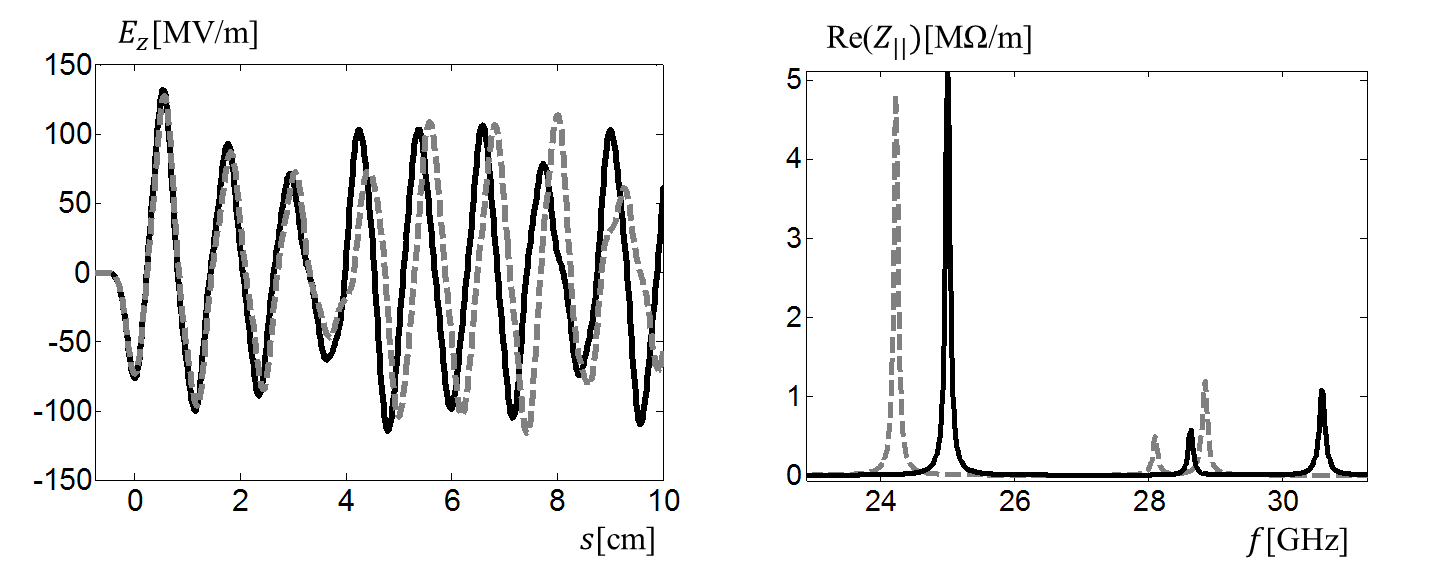}
	\caption{The longitudinal electric field component and the real part of the longitudinal impedance for anisotropic (solid black line) and isotropic (dashed gray line) rectangular structures.}\label{FigNE3}
\end{figure}

For the same aperture size the cylindrical geometry allows to obtain the highest accelerating gradients. Due to  technological difficulties in preparing cylindrical structures with stringent requirements to tolerances the rectangular structures are considered as well. As a next example we consider a Gaussian relativistic electron bunch with parameters of the Argonne wakefield accelerator in the sapphire-based rectangular accelerating structure~\cite{Kan10,Sheinman}. The rectangular structure has width $2w = 11$mm, the anisotropic layer starts at $a_0= 1.5$ mm and is closed by PEC at $a_1 = 2.39$ mm. The permittivities along main axes are: $\epsilon_x=\epsilon_z=9.4\epsilon_0,\epsilon_y=11.5\epsilon_0$. It corresponds to a frequency of 25.0 GHz of the accelerating mode of
the structure. For comparison a waveguide with isotropic
dielectric filling with $\epsilon=10.45\epsilon_0$ corresponds to the base frequency of 24.23 GHz. The electron bunch with energy $15$ MeV, charge $100$ nC and bunch length $\sigma_z = 1.5$mm is considered. The dependence of the longitudinal electric field component $E_z$ at the symmetry axis produced by the bunch on the distance $s = v t-z$ behind it is shown in Fig.~\ref{FigNE3}. The solid line corresponds to anisotropic sapphire, the dashed line corresponds to isotropic filling. The wave number $k$ was sampled from $1$ $m^{-1}$ to $20e4$ $m^{-1}$ with step 0.2 and we have calculated 5 the lowest odd Fourier harmonics in Eq.~(\ref{EqImpRectWG}). At this example we used a small conductivity $\kappa=0.05$ S/m to resolve the real part of the impedance. The data in Fig.~\ref{FigNE3} agree with the results published in~\cite{Sheinman}. A frequency shift with a little influence on the wake field amplitudes can be seen. 

The execution times of different methods discussed in this paper for the rectangular example are shown in Table~\ref{Table01}. It can be seen again that for the same accuracy the combined method requires less computational time as compared to a fully finite-difference one.
%
\section*{Acknowledgements}
%

The author thanks K.L.F. Bane, M. Dohlus, F. Lemery and  G. Stupakov for helpful discussions.

\end{document}